\documentstyle[nato,epsf]{crckapb}

\def\go{\mathrel{\raise.3ex\hbox{$>$}\mkern-14mu\lower0.6ex\hbox{$\sim$}}}
\def\lo{\mathrel{\raise.3ex\hbox{$<$}\mkern-14mu\lower0.6ex\hbox{$\sim$}}}

\begin{opening}

\title{The Final Fate of Coalescing Compact Binaries:\protect\\
From Black Hole to Planet Formation}

\author{Frederic A.~Rasio}

\institute{Department of Physics,\\
 Massachusetts Institute of Technology,\\
 Cambridge, MA 02139, USA}

\end{opening}

\runningtitle{Coalescing Compact Binaries}

\begin{document}

\begin{abstract}
Coalescing compact binaries are thought to be involved in a wide variety 
of astrophysical phenomena. In particular, they are important sources of 
gravitational radiation for both ground-based and space-based laser-interferometer
detectors, and they may be sources of supernova explosions or gamma-ray bursts. 
Mergers of two white dwarfs may produce neutron stars with peculiar
properties, including perhaps millisecond radio pulsars sometimes
accompanied by planets (as observed in PSR 1257+12). According to a widely
held belief, the coalescence of two neutron stars should produce a rapidly 
rotating black hole surrounded by an accretion disk or torus, but this is by 
no means certain. This review paper focuses on the final hydrodynamic 
coalescence and merger of double neutron stars and double white dwarfs, 
and addresses the question of the nature of the final merger products.
\end{abstract}

\section{Introduction}

The coalescence and merging of two compact stars into a single object
is a very common end-point of close binary evolution.  
Dissipation mechanisms such as friction in common envelopes, tidal dissipation,
or the emission of gravitational radiation, are always present and cause 
the orbits of close binary systems to decay. 
This review will concentrate on the coalescence of compact binaries
containing either two neutron stars (hereafter NS) or two white dwarfs (WD). 

\subsection{Double Neutron Stars}

Many theoretical models of gamma-ray bursts (GRBs) 
rely on coalescing NS 
binaries to provide the energy of GRBs at
cosmological distances (e.g., Eichler et al.\ 1989; Narayan,
Paczy\'nski, \& Piran 1992; M\'esz\'aros \& Rees 1992;
for recent reviews see M\'esz\'aros 1999 and Piran 1999). 
The close spatial association of some GRB afterglows 
with faint galaxies at high redshifts may not be inconsistent
with a NS binary merger origin, in spite of the large recoil
velocities acquired by NS binaries at birth (Bloom,
Sigurdsson, \& Pols 1999; but see also Bulik \& Belczynski 2000).
Currently the most popular models all assume that the coalescence of two
NS leads
to the formation of a rapidly rotating black hole (BH) 
surrounded by a torus of debris. 
Energy can then be extracted either from the rotation of the 
Kerr BH or from
the material in the torus so that, with sufficient beaming, the
gamma-ray fluxes observed from even the most distant GRBs can be
explained (M\'esz\'aros, Rees, \& Wijers 1999). However, it is important to
understand the hydrodynamic processes taking place during the final 
coalescence before making assumptions about its outcome. In particular,
as will be argued below (\S2.2), it is not clear that the coalescence of
two $1.4\,M_\odot$ NS forms an object that will collapse to a BH
on a dynamical timescale,
and it is not certain either that a significant amount of 
matter will be ejected
during the merger to form an outer torus around the central object
(Faber \& Rasio 2000).

Coalescing NS binaries are also important sources of gravitational
waves that may be directly detectable by the large laser interferometers 
currently under construction,
such as LIGO (Abramovici et al.\ 1992; see Barish \& Weiss 1999 for a
recent pedagogical introduction) and VIRGO (Bradaschia et al.\ 1990). 
In addition to providing a major new confirmation of
Einstein's theory of general relativity (GR), including the first direct
proof of the existence of black holes (see, e.g., Flanagan \& Hughes 1998;
Lipunov, Postnov, \& Prokhorov 1997), the detection of gravitational
waves from coalescing binaries at cosmological distances could provide 
accurate independent measurements of the Hubble constant
and mean density of the Universe (Schutz 1986; Chernoff \& Finn 1993; 
Markovi\'c 1993). 
Expected rates of NS binary coalescence in the Universe, 
as well as expected event rates in laser interferometers, have 
now been calculated by many groups. Although there is some disparity 
between various published results, the estimated rates are generally 
encouraging (see Kalogera 2000 for a recent review).

Many calculations of gravitational wave emission from coalescing binaries 
have focused on the waveforms emitted during the last few thousand orbits, 
as the frequency sweeps upward from $\sim10\,$Hz to $\sim300\,$Hz.
The waveforms in this frequency range, where the sensitivity of
ground-based interferometers 
is highest, can be calculated very accurately by 
performing high-order post-Newtonian (PN)
expansions of the equations of 
motion for two {\it point masses\/} (see, e.g., Owen \& Sathyaprakash 1999
and references therein). However, at the end of the inspiral, 
when the binary separation becomes comparable 
to the stellar radii (and the frequency is $\go1\,$kHz), 
hydrodynamics becomes important and the character 
of the waveforms must change. 
Special purpose narrow-band detectors that can sweep up frequency in real 
time will be used to try to catch the last $\sim10$ cycles of the gravitational
waves during the final coalescence
(Meers 1988; Strain \& Meers 1991). These ``dual recycling''
techniques are being tested right now on the German-British interferometer
GEO 600 (Danzmann 1998). In this terminal phase of the coalescence,
when the two stars merge together into a single object, 
the waveforms contain information not just about the 
effects of GR, but also about the interior structure 
of a NS and the nuclear equation of state 
(EOS) at high density. 
Extracting this information from observed waveforms, 
however, requires detailed theoretical knowledge about all relevant
hydrodynamic processes. 
If the NS merger is followed by the formation 
of a BH, the corresponding gravitational radiation waveforms will also 
provide direct information on the dynamics of rotating core collapse
and the BH ``ringdown'' (see, e.g., Flanagan \& Hughes 1998).

\subsection{Double White Dwarfs}

Coalescing WD binaries have long been discussed as possible progenitors
of Type~Ia supernovae (Iben \& Tutukov 1984; Webbink 1984; Paczy\'nski 1985;
see Branch et al.\ 1995 for a recent review). To produce a supernova,
the total mass of the system must be above the Chandrasekhar mass. Given
evolutionary considerations, this requires two C-O or O-Ne-Mg WD.
Yungelson et al.\ (1994) showed that the expected merger rate for close WD
pairs with total mass exceeding the Chandrasekhar mass is consistent
with the rate of Type~Ia supernovae deduced from observations.
Alternatively, a massive enough merger may collapse to form a rapidly
rotating NS (Nomoto \& Iben 1985; Colgate 1990).
Chen \& Leonard (1993) speculated that most
millisecond pulsars in globular clusters might have formed in this way.
In some cases planets may also form in the disk of material ejected during
the coalescence and left in orbit around the central pulsar 
(Podsiadlowski, Pringle, \& Rees 1991). Indeed the very
first extrasolar planets were discovered
in orbit around a millisecond pulsar, PSR B1257$+$12 (Wolszczan \& Frail 1992).
A merger of two magnetized WD might lead to the formation of  
a NS with extremely high magnetic field, and this scenario has been 
proposed as a source of GRBs (Usov 1992). 

Close WD binaries are expected to be extremely abundant in our Galaxy,
even though their direct detection remains very challenging (Han 1998;
Saffer, Livio, \& Yungelson 1999).
Iben \& Tutukov (1984, 1986) predicted that $\sim20$\% of all binary stars 
produce close WD pairs at the end of their stellar evolution.
More recently, theoretical estimates of the double WD formation rate
in the Galaxy have converged to a value 
$\simeq0.1\,{\rm yr}^{-1}$, with an uncertainty that may be only a factor of 
two (Han 1998; Kalogera 2000). 
The most common systems should be those containing two low-mass helium WD.
Their final coalescence can produce an object massive enough
to start helium burning. Bailyn (1993 and references therein) and others 
have suggested that some
``extreme horizontal branch'' stars in globular clusters 
may be such helium-burning stars formed by the coalescence of two WD.
Planets in orbit around a massive WD may also form following
the binary coalescence (Livio, Pringle, \& Saffer 1992).

Coalescing WD binaries are also important sources of low-frequency 
gravitational waves
that should be easily detectable by future space-based laser interferometers.
The currently planned LISA (Laser Interferometer Space Antenna; see Folkner 1998)
should have an
extremely high sensitivity (down to a characteristic strain $h\sim10^{-23}$)
to sources with frequencies in the range $\sim10^{-4}\,-\,1\,$Hz.
Han (1998) estimated a WD merger rate 
$\sim0.03\,{\rm yr}^{-1}$ in our own Galaxy. Individual coalescing systems 
and mergers may be detectable in the frequency range $\sim10$--$100\,$mHz. 
In addition, the total number ($\sim10^4$) of close WD binaries 
in our Galaxy emitting at lower frequencies $\sim0.1$--$10\,$mHz (the emission 
lasting for $\sim10^2$--$10^4\,$yr before final merging) should 
provide a continuum background signal of amplitude 
$h\sim10^{-20}$--$10^{-21}$ (Hils et al.\ 1990).
The detection of the final burst of gravitational waves emitted during 
an actual merger would provide a unique opportunity to observe in ``real time'' 
the hydrodynamic interaction between the two degenerate stars, possibly followed
immediately by a supernova explosion, nuclear outburst, or some other type of 
electromagnetic signal.

\section{Coalescing Binary Neutron Stars}

\subsection{Hydrodynamics of Neutron Star Mergers}

The final hydrodynamic merger of two NS is driven by a combination
of relativistic and fluid effects. Even in Newtonian gravity,
an innermost stable circular orbit (ISCO) is imposed by
{\it global hydrodynamic instabilities\/}, which can drive 
a close binary system to rapid coalescence once the tidal interaction 
between the two stars becomes sufficiently strong.
The existence of these global instabilities 
for close binary equilibrium configurations containing a compressible fluid, 
and their particular importance for binary NS systems, 
were demonstrated for the first time by 
Rasio \& Shapiro (1992, 1994, 1995; hereafter RS1--3) 
using numerical hydrodynamic calculations.
These instabilities can also be studied using analytic methods.
The classical analytic work for close binaries containing an
incompressible fluid (e.g., Chandrasekhar 1969) was
extended to compressible fluids in the work of Lai, Rasio, \& Shapiro 
(1993a,b, 1994a,b,c, hereafter LRS1--5).
This analytic study confirmed the existence of dynamical 
instabilities for sufficiently close binaries.
Although these simplified analytic studies can give much physical
insight into difficult questions of global fluid instabilities, 
fully numerical calculations remain essential for establishing
the stability limits of close binaries accurately and for following 
the nonlinear evolution of unstable systems all the way to complete 
coalescence. 

A number of different groups have now performed such calculations, using
a variety of numerical methods and focusing on different aspects of the
 problem. Nakamura and collaborators (see Nakamura \& Oohara 1998 and 
references therein)
were the first to perform 3D hydrodynamic calculations of binary 
NS coalescence, using a traditional Eulerian finite-difference code. 
Instead, RS used the 
Lagrangian method SPH (Smoothed Particle Hydrodynamics). They focused
on determining the ISCO for initial binary models in strict
hydrostatic equilibrium and calculating the emission of gravitational waves
from the coalescence of unstable binaries. Many of the results of RS were
later independently confirmed by New \& Tohline (1997) and Swesty,
Wang, \& Calder (1999), who used completely
different numerical methods but also focused on stability questions, and 
by Zhuge, Centrella, \& McMillan (1994, 1996), who also 
used SPH. Zhuge et al.\ (1996) also explored in detail the dependence of
the gravitational wave signals on the initial NS spins. 
Davies et al.\ (1994) and Ruffert et al.\ (1996, 1997) have
incorporated a treatment of the nuclear physics in their hydrodynamic
calculations (done using SPH and PPM codes, respectively), motivated
by models of GRBs at cosmological distances.
All these calculations were performed in {\it Newtonian gravity\/}, with
some of the more recent studies adding an approximate treatment of
energy and angular momentum dissipation through the gravitational 
radiation reaction (e.g., Janka et al.\ 1999; Rosswog et al.\ 1999),
or even a full treatment of PN gravity to lowest order (Ayal et al.\ 2000;
Faber \& Rasio 2000).

All recent hydrodynamic calculations agree on
the basic qualitative picture that emerges for the final coalescence
(see Fig.~1). As the ISCO is approached, the secular orbital
decay driven by gravitational wave emission is dramatically accelerated
(see also LRS2, LRS3).
The two stars then plunge rapidly toward each other, and merge together 
into a single object in just a few rotation periods. In the corotating 
frame of the binary, the relative radial velocity of the two stars always 
remains very subsonic, so that the evolution is nearly adiabatic.
This is in sharp contrast to the case of a head-on collision between
two stars on a free-fall, radial orbit, where
shock heating is very important for the dynamics (RS1; Shapiro 1998).
Here the stars are constantly being held back by a (slowly receding)
centrifugal barrier, and the merging, although dynamical, is much more gentle. 
After typically $1-2$ orbital periods following first contact,
 the innermost cores of the 
two stars have merged and 
a secondary instability occurs: {\it mass shedding\/} 
sets in rather abruptly. Material (typically $\sim10$\% of the total mass) 
is ejected through the outer Lagrange
points of the effective potential and spirals out rapidly.
In the final stage, the spiral arms widen and merge together, 
forming a nearly axisymmetric thick disk or torus around the inner, 
maximally rotating dense core. 

\begin{figure}
\epsfxsize=5.truein
\epsfbox{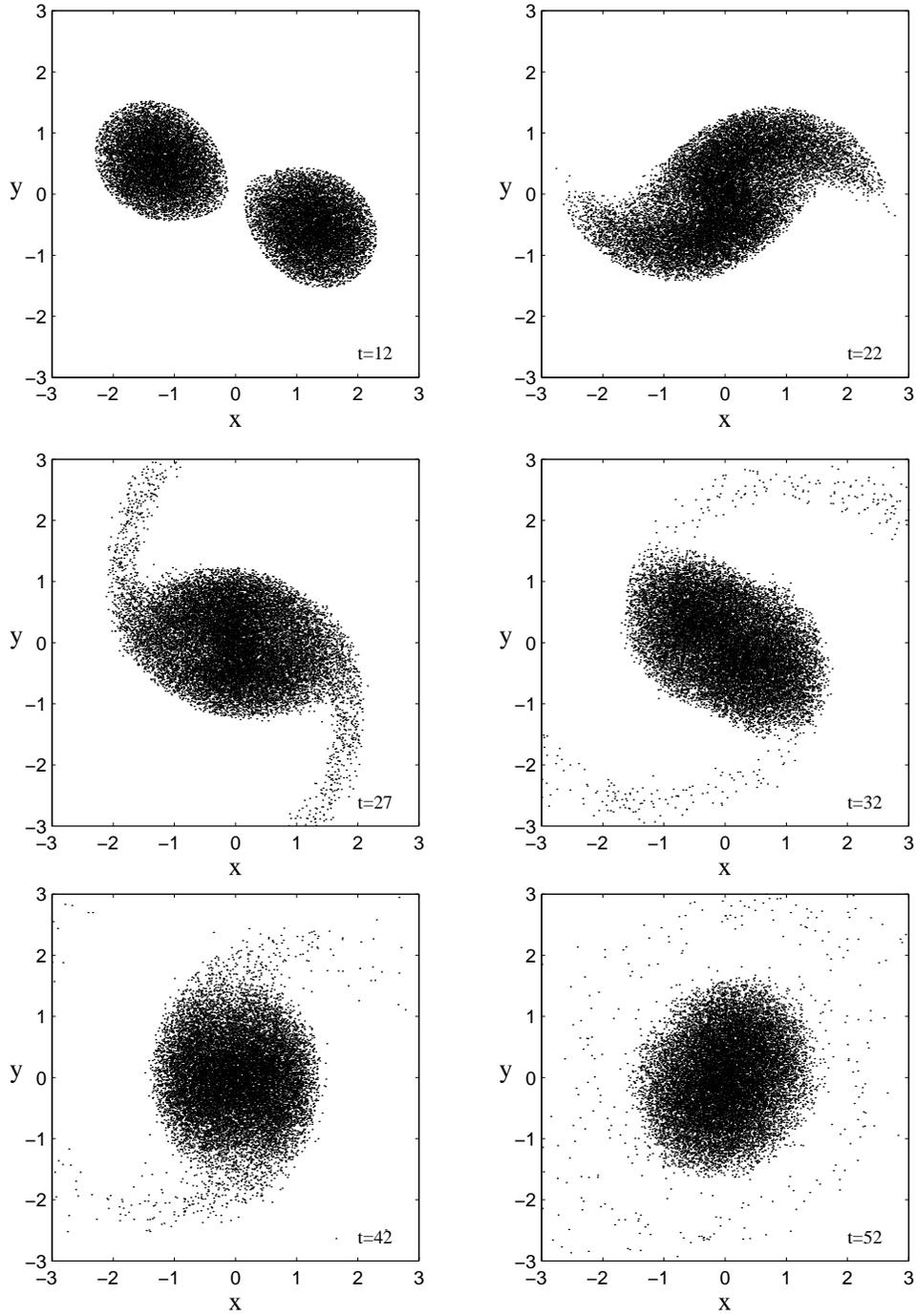}
\caption{Post-Newtonian SPH calculation of the coalescence of two identical
neutron stars modeled as simple $\Gamma=3$ polytropes. Projections of a subset
of all SPH particles onto the orbital ($x-y$) plane are shown at various times.
Units are such that $G=M=R=1$ where $M$ and $R$ are the mass and radius
of each star initially. The orbital rotation is counter-clockwise. From
Faber \& Rasio (2000).}
\end{figure}

In GR, strong-field gravity between the masses in
a binary system is alone sufficient to drive a close circular 
orbit unstable. In close NS binaries, GR effects combine nonlinearly
with Newtonian tidal effects so that the ISCO is encountered
at larger binary separations and lower orbital frequency than 
predicted by Newtonian hydrodynamics alone, or GR alone for two point
masses. The combined effects
of relativity and hydrodynamics on the stability of close compact
binaries have only very recently begun to be studied,
using both analytic approximations
(basically, PN generalizations of LRS; see, e.g., 
Lai \& Wiseman 1997; Lombardi, Rasio, \& Shapiro 1997; 
Shibata \& Taniguchi 1997), as well as numerical 
calculations in 3D incorporating simplified treatments of 
relativistic effects 
(e.g., Baumgarte et al.\ 1998; Marronetti, Mathews \& Wilson 1998;  
Wang, Swesty, \& Calder 1998; Faber \& Rasio 2000).

Several groups have been working on a fully general relativistic
calculation of the final coalescence, combining the techniques of 
numerical relativity and numerical hydrodynamics in 3D
(Baumgarte, Hughes, \& Shapiro 1999;
Landry \& Teukolsky 2000; Seidel 1998; Shibata \& Uryu 2000). 
However this work is still in its infancy, and only very preliminary results
of test calculations have been reported so far.

\subsection{Black Hole Formation}

The final fate of a NS--NS merger depends crucially on the NS EOS,
and on the extraction of angular momentum from the system during the 
final merger. For a stiff NS EOS, it is by no means
certain that the core of the final merged configuration will collapse
on a dynamical timescale to form a BH. One reason is that the Kerr
parameter $J/M^2$ of the core may exceed unity for extremely stiff
EOS (Baumgarte et al.\ 1998), although Newtonian and PN 
hydrodynamic calculations suggest that this is never the case
(see, e.g., Faber \& Rasio 2000). 
More importantly, the rapidly rotating core may in fact be 
dynamically stable. 

Take the obvious example of a system containing two 
identical $1.35\,M_\odot$ NS. The total baryonic mass of the system
for a stiff NS EOS is then about $3\,M_\odot$. Almost independent of 
the spins of the NS, all hydrodynamic calculations suggest that about
$10\%$ of this mass will be ejected into the outer torus, leaving at
the center a {\it maximally rotating\/} object with baryonic mass 
$\simeq2.7\,M_\odot$ (Any hydrodynamic merger process that leads to mass
shedding will produce a maximally rotating object since the system will
have ejected just enough mass and angular momentum to reach its new,
stable quasi-equilibrium state). Most stiff NS EOS (including the
well-known ``AU'' and ``UU'' EOS of Wiringa et al.\ 1988; see Akmal et al.\
1998 for a recent update) allow stable,
maximally rotating NS with baryonic masses exceeding
$3\,M_\odot$ (Cook, Shapiro, \& Teukolsky 1994), i.e., well above the mass
of the final merger core. Differential rotation (not taken into account in the
calculations of Cook et al.\ 1994) can further increase this maximum stable 
mass very significantly (see Baumgarte, Shapiro, \& Shibata 2000).
Thus the hydrodynamic merger of two NS with stiff EOS and realistic
masses is not expected to produce a BH. This expectation is confirmed by
the preliminary full-GR calculations of Shibata \& Uryu (2000), for
polytropes with $\Gamma=2$, which indicate collapse to a BH only
when the two NS are initially very close to the maximum stable
mass. 

For {\it slowly rotating\/} stars, the same stiff NS EOS give
maximum stable baryonic masses in the range $2.5-3\,M_\odot$, which may
or may not exceed the total merger core mass. Therefore, collapse to
a BH could still occur on a timescale longer than the dynamical timescale,
following a significant loss of angular momentum.
Indeed, processes such as 
electromagnetic radiation, neutrino emission, and the development of
various secular instabilities (e.g., r-modes), which may lead to angular
momentum losses, take place on timescales much longer than the dynamical
timescale (see, e.g., Baumgarte \& Shapiro 1998, who show that
neutrino emission is probably negligible). These processes are
therefore decoupled from the hydrodynamics of the coalescence.
Unfortunately their
study is plagued by many fundamental uncertainties in the microphysics.

\subsection{The Importance of the Neutron Star Spins}

The question of the final fate of the merger could also depend crucially
on the NS spins and on the 
evolution of the fluid vorticity during the final coalescence.
Close NS binaries  are likely to be {\it nonsynchronized\/}. Indeed,
the tidal synchronization time 
is almost certainly much longer than the orbital decay
time (Kochanek 1992; Bildsten \& Cutler 1992).
For NS binaries that are far from synchronized,
the final coalescence involves
some new, complex hydrodynamic processes (Rasio \& Shapiro 1999).

Consider for example the case of an irrotational system (containing
two nonspinning stars at large separation; see LRS3).
Because the two stars appear to be counter-spinning in the corotating
frame of the binary, a {\it vortex sheet\/}  (where the tangential velocity
jumps discontinuously by $\Delta v\sim 0.1\,c$) appears when the stellar 
surfaces come into contact.
Such a vortex sheet is Kelvin-Helmholtz unstable on all 
wavelengths and the hydrodynamics is therefore  extremely
difficult to model accurately given the limited spatial
resolution of 3D calculations.
The breaking of the vortex sheet generates some turbulent
viscosity so that the final configuration may no longer be
irrotational. In numerical simulations, however, vorticity is
quickly generated through spurious shear viscosity, and the 
merger remnant is observed to evolve rapidly (in just a few
rotation periods) toward uniform rotation.

The final fate of the merger could be affected drastically by these
processes. In particular, the shear flow inside the merging stars 
(which supports a highly triaxial shape; see Rasio \& Shapiro 1999) may
in reality persist long enough to allow a large fraction of the total
angular momentum
in the system to be radiated away in gravitational waves during the
hydrodynamic phase of the coalescence. In this case
the final merged core may resemble a Dedekind ellipsoid, i.e., it will have
a triaxial shape supported entirely by internal fluid motions, but with
a stationary shape in the inertial frame (so that it no longer
radiates gravitational waves). 
This state will be reached on the gravitational radiation reaction
timescale, which is no more than a few tens of rotation periods.
On the (much longer) {\it viscous timescale\/}, the core will
then evolve to a uniform, slowly rotating state and will probably collapse 
to a BH.
In contrast, in all 3D numerical simulations performed to date,
the shear is quickly dissipated, so that gravitational radiation
never gets a chance to extract more than a small fraction ($\lo10$\%)
of the angular momentum, and the final core appears to be a uniform,
maximally rotating object (stable to collapse) exactly as in calculations starting
from synchronized binaries. However this behavior is most likely
an artefact of the large spurious shear viscosity present in the
3D simulations. 

In addition to their obvious significance for gravitational wave emission,
these issues are also of great importance for models 
of GRBs that depend on energy extraction from a torus of material around
the central BH. Indeed, if a large fraction of the total angular momentum
is removed by the gravitational waves, rotationally-induced mass shedding 
may not occur at all during the merger, eventually
leaving a BH with no surrounding 
matter, and no way of extracting energy from the system.
Note also that, even without any additional loss of angular momentum
through gravitational radiation, PN effects tend to reduce drastically the
amount of matter ejected during the merger (Faber \& Rasio 2000).

\section{Coalescing White Dwarf Binaries}

\subsection{Hydrodynamics of White Dwarf Mergers}

The results of RS3 for polytropes with $\Gamma=5/3$ show that
hydrodynamics also plays an important role in the coalescence of two
WD, either because of dynamical instabilities of the
equilibrium configuration, or following the onset of dynamically unstable
mass transfer. Systems with mass ratios $q\approx1$ must evolve into deep contact
before they become dynamically unstable and merge. Instead, equilibrium
configurations for binaries with $q$ 
sufficiently far from unity never become dynamically unstable. However,
once these binaries reach their Roche limit, dynamically unstable mass
transfer occurs and the less massive star is completely disrupted after
a small number ($<10$) of orbital periods (see also Benz et al.\ 1990). In both
cases, the final merged configuration is an axisymmetric, rapidly rotating 
object with a core -- thick disk structure similar to that obtained for coalescing 
NS (RS2, RS3; see also Mochkovitch \& Livio 1989).

\subsection{The Final Fate: Collapse to a Neutron Star? Planets?}

For two massive enough WD, the merger product may be well above the 
Chandrasekhar mass $M_{Ch}$.
The object may therefore explode as a (Type~Ia) supernova, or perhaps collapse 
to a NS. The rapid rotation and possibly high mass (up to $\sim2M_{Ch}$) 
of the object must be taken into account for determining its final fate.
Unfortunately, rapid rotation and the possibility of starting from an object
well above the Chandrasekhar limit have not been taken into account in 
most previous theoretical calculations of ``accretion-induced collapse''
 (AIC), which consider a 
nonrotating WD just below the Chandrasekhar limit 
accreting matter slowly and quasi-spherically (e.g., Canal et al.\ 1990; 
Nomoto \& Kondo 1991; see Fryer et al.\ 1999 for a
recent 2-D SPH calculation including rotation). 
Under these assumptions it is found that collapse to a NS
is possible only for a narrow range of initial conditions.
In most cases, a supernova explosion follows the ignition of the
nuclear fuel in the degenerate core.
However, the fate of a much more massive object with substantial 
rotational support and large deviations from spherical symmetry 
(as would be formed by dynamical coalescence) may be very different.

If a NS does indeed form, and later accretes some of the 
material ejected during the coalescence, a millisecond radio pulsar 
may emerge. Planets around this millisecond pulsar may be formed
at large distances $\sim1\,$AU following the viscous evolution of the remaining
material in the outer disk 
(Podsiadlowski, Pringle \& Rees 1991; Phinney \& Hansen 1993).
This is one of the possible formation scenarios for the extraordinary
planetary system discovered around the millisecond pulsar PSR~B1257$+$12
(see Wolszczan 1999 for a recent update; Podsiadlowski 1993 for
alternative planet formation scenarios). 
This system contains three 
confirmed Earth-mass planets in quasi-circular orbits 
(Wolszczan \& Frail 1992; Wolszczan 1994).  
The planets have masses of $0.015/\sin i_1\,\rm M_\oplus$, 
$3.4/\sin i_2\,\rm M_\oplus$, and $2.8/\sin i_3\,\rm M_\oplus$, where $i_1$, 
$i_2$ and $i_3$ are the inclinations of the orbits with respect to the 
line of sight, 
and are at distances of 0.19\,AU, 0.36\,AU, and 0.47\,AU, respectively, 
from the pulsar. In addition, the unusually large second and third frequency 
derivatives of the pulsar suggest the existence of a fourth, more distant 
and massive planet in the system (Wolszczan 1999).
The simplest interpretation of the 
present best-fit values of the frequency derivatives implies 
a mass of about $100/\sin i_4 \,\rm M_\oplus$ (i.e., comparable to 
Saturn's mass) for the fourth planet, at a distance of about
$38\,\rm AU$ (i.e., comparable to Pluto's distance from the Sun), and
with a period of about $170\,\rm yr$ in a circular, coplanar orbit 
(Wolszczan 1996; Joshi \& Rasio 1997). However, if, as may well be the case,
the first pulse frequency derivative is not entirely acceleration-induced, 
then the fourth planet can have a wide range of masses
(Joshi \& Rasio 1997).
In particular, it can have a mass comparable to that of 
Mars (at a distance of $9\,$AU), Uranus (at a distance of $25\,$AU) or 
Neptune (at a distance of $26\,$AU).
The presence of this fourth planet, if confirmed,
would place strong additional constraints on possible formation
scenarios, as both the minimum mass and minimum angular momentum
required in the protoplanetary disk would increase considerably
(see Phinney \& Hansen 1993 for a general discussion).

\section*{Acknowledgements}

This work was supported by NSF Grant AST-9618116, NASA ATP Grant NAG5-8460, 
and by an Alfred P.\ Sloan Research Fellowship. Our computational
work is supported by the National Computational Science Alliance.

\end{document}